\documentclass[conference]{IEEEtran}
\IEEEoverridecommandlockouts
\usepackage{amsfonts}
\usepackage{algorithm} 
\usepackage{algpseudocode} 
\usepackage{algorithm, setspace}
\usepackage{textcomp}
\usepackage{xcolor}
\usepackage{cite}
\usepackage{graphicx,color,epsfig,rotating}
\usepackage{amsfonts,amsmath,amssymb,bbm}
\usepackage{caption}
\usepackage{subcaption}
\usepackage{cite}
\usepackage{placeins}
\usepackage{graphicx}
\usepackage[latin1]{inputenc}
\usepackage{multirow}
\usepackage{stfloats}
\usepackage{tabularx} 
\usepackage{booktabs} 
\usepackage{url}
\usepackage{bm}
\usepackage{soul}
\usepackage[normalem]{ulem}

\long\def\comment#1{}

\newfont{\bbb}{msbm10 scaled 700}

\newfont{\bb}{msbm10 scaled 1100}

\newcommand{\EE}{\mbox{\bb E}}

\newcommand{\sfu}{{\textrm{UE}}}
\newcommand{\sfb}{{\textrm{BS}}}

\newcommand{\Ac}{{\cal A}}

\newcommand{\Dc}{{\cal D}}

\newcommand{\Ic}{{\cal I}}

\newcommand{\Kc}{{\cal K}}

\newcommand{\Mc}{{\cal M}}

\newcommand{\Oc}{{\cal O}}
\newcommand{\Pc}{{\cal P}}

\newcommand{\Sc}{{\cal S}}

\renewcommand{\arg}{{\hbox{arg}}}

\newcommand{\eqdef}{\stackrel{\Delta}{=}}

\newcommand{\be}{\begin{equation}}
\newcommand{\ee}{\end{equation}}
\newcommand{\bea}{\begin{eqnarray}}
\newcommand{\eea}{\end{eqnarray}}

\newtheorem{remark}{Remark}
\makeatletter
\newcommand{\subalign}[1]{
  \vcenter{%
    \Let@ \restore@math@cr \default@tag
    \baselineskip\fontdimen10 \scriptfont\tw@
    \advance\baselineskip\fontdimen12 \scriptfont\tw@
    \lineskip\thr@@\fontdimen8 \scriptfont\thr@@
    \lineskiplimit\lineskip
    \ialign{\hfil$\m@th\scriptstyle##$&$\m@th\scriptstyle{}##$\crcr
      #1\crcr
    }%
  }
}

\newcommand{\thickhline}{%
    \noalign {\ifnum 0=`}\fi \hrule height 1pt
    \futurelet \reserved@a \@xhline
}
\newcolumntype{"}{@{\hskip\tabcolsep\vrule width 1pt\hskip\tabcolsep}}
\makeatother

\ifodd 1

\else

\fi

\ifodd 1

\else

\fi

\def\BibTeX{{\rm B\kern-.05em{\sc i\kern-.025em b}\kern-.08em
    T\kern-.1667em\lower.7ex\hbox{E}\kern-.125emX}}

\title{A Q-Learning-based Approach for Distributed Beam Scheduling in mmWave Networks}

\makeatletter
\newcommand{\linebreakand}{%
  \end{@IEEEauthorhalign}
  \hfill\mbox{}\par
  \mbox{}\hfill\begin{@IEEEauthorhalign}
}
\makeatother

\author{\IEEEauthorblockN{Xiang Zhang\IEEEauthorrefmark{1}, Shamik Sarkar\IEEEauthorrefmark{2},  Arupjyoti Bhuyan\IEEEauthorrefmark{3}, Sneha Kumar Kasera\IEEEauthorrefmark{2}\IEEEauthorrefmark{1} and Mingyue Ji\IEEEauthorrefmark{1}\IEEEauthorrefmark{2}}
\IEEEauthorblockA{Department of Electrical and Computer Engineering, University of Utah\IEEEauthorrefmark{1}\\
School of Computing, University of Utah \IEEEauthorrefmark{2}\\
Idaho National Laboratory\IEEEauthorrefmark{3}\\
Email: \IEEEauthorrefmark{1}\{xiang.zhang, mingyue.ji\}@utah.edu,\IEEEauthorrefmark{2}shamik.sarkar@utah.edu, \IEEEauthorrefmark{2}kasera@cs.utah.edu, \IEEEauthorrefmark{3}arupjyoti.bhuyan@inl.gov}}

\begin{document}
\newcolumntype{A}{>{\centering}p{0.025\textwidth}}
\newcolumntype{B}{p{0.2\textwidth}}

\maketitle
\begin{abstract}
We consider the problem of distributed downlink beam scheduling and power allocation for millimeter-Wave (mmWave) cellular networks where multiple base stations (BSs) belonging to different service operators share the same unlicensed spectrum with no central coordination or 
cooperation among them. 
Our goal is to design efficient distributed beam scheduling and power allocation algorithms such that the network-level payoff, defined as the weighted sum of the total throughput and a power penalization term, can be maximized. To this end, we propose a distributed scheduling approach to power allocation and adaptation for efficient interference management over the shared spectrum by modeling each BS as an independent Q-learning agent. As a baseline, we compare the proposed approach to the state-of-the-art non-cooperative game-based approach which was previously developed for the same problem. We conduct extensive experiments under various scenarios to verify the effect of multiple factors on the performance of both approaches. Experiment results show that the proposed approach adapts well to different interference situations by learning from experience  and can achieve 
higher payoff than the game-based approach.
The proposed approach can also be 
integrated into our previously developed Lyapunov stochastic optimization framework for the purpose of network utility maximization with optimality guarantee. 
As a result,  
the weights in the payoff function can be automatically and optimally determined by the virtual queue values from the sub-problems derived from the Lyapunov optimization framework.
\end{abstract}

\begin{IEEEkeywords}
mmWave, distributed scheduling, reinforcement learning, Q-learning, optimality
\end{IEEEkeywords}

\section{Introduction}
The proliferation of mmWave frequencies in 5G cellular networks has increased wireless bandwidth by orders of magnitude and has also enhanced spectrum availability. Spectrum sharing enables the secondary utilization of additional unlicensed  or shared spectrum that is available by allowing concurrent beam-based transmission and has the potential to largely enhance the the system-level throughput performance \cite{boccardi2016spectrum,gupta2016feasibility,jorswieck2014spectrum}. However, highly directional transmission may also present a severe interference condition, which is even worsened by the dense population of access points and user equipment (UEs), if there is no proper coordination of the beams.   To handle interference and improve system throughput, two major paradigms -- centralized and distributed, are considered in the literature. In general, centralized approaches can be effective but are usually limited by high complexity and limited scalability especially for large networks.  

Distributed approaches \cite{980097,pang2010design,ning2019interference,sarkar2021uncoordinated,candogan2010near,zhang2020non,galindo2010distributed,van2012power,kar2013distributed
,ghadimi2017reinforcement }, on the other hand, are scalable and have the advantage of improving system security by removing any central point of attack as well as any falsification of reports of spectrum use to the centralized server. 
Some existing distributed approaches~\cite{980097,pang2010design,ning2019interference,sarkar2021uncoordinated,candogan2010near,zhang2020non}, including our own [9], have proposed the use of game theory, for beam scheduling.
In [9], combined with the Lyapunonv optimization framework, we proposed a distributed game-based beam scheduling for mmWave networks with non-cooperative operators. 

In this paper, we propose an alternate approach that uses Q-learning for distributed beam scheduling as well as for power allocation for mmWave networks with non-cooperative operators. Our main contributions are two-fold. First, we present a general framework for dynamic spectrum sharing for the purpose of optimizing a network- level payoff function, which is defined as the sum throughput penalized by power consumption. The weights in the payoff function can be tuned to find a desirable trade-off between throughput maximization and power consumption. This formulation can work for various different beam scheduling methods and therefore, provides a unified framework for performance evaluation and comparison of these methods. Second, under the proposed payoff optimization framework, we apply classical Q-learning due to its simplicity and yet promising performance. We propose a learning-based power allocation algorithm by modeling each base station (BS) as an independent Q-learning agent that interacts with the radio environment determined by the joint actions of all BSs and channel uncertainty. We compare the proposed approach with our, state-of-the-art, non-cooperative game-based approach [9] and demonstrate that the our learning approach adapts well to different interference situations and achieves higher network-level payoff than the game-based approach with a greedy nature. We conduct simulation-based experiments under various interference scenarios and the results show that our approach can achieve 23\% to 80\% more payoff than the game-based approach under practical network settings, and the performance gain is more prominent in the relatively low SINR regime. In addition, our approach can be integrated seamlessly into a general network utility maximization framework by using the Lyapunov stochastic optimization proposed in [9]. In this case, the weights in the payoff function can be automatically and optimally determined by the virtual queues derived from the Lyapunov optimization. Further experiments show that the proposed approach can also achieve 7\% to 23\% more utility than the game-based approach. This performance gain is significant as the utility is defined as an increasing concave function of the average throughput which has a diminishing marginal utility.


\noindent {\bf Why Reinforcement Learning:} In general, reinforcement learning-based methods have the advantage of being adaptive to different interference conditions by learning from experience, i.e., past interaction with the environment, the quality of each decision made indicated by the corresponding reward. In addition, by actively exploring non-greedy actions, there is a higher chance of finding the optimal actions in the long run. In contrast, the game-based methods are greedy by nature -- regardless of the interference, each BS will always choose an action that maximizes its payoff in the current step. This greedy nature prevents the BSs from exploring non-greedy actions or adapting their decisions to different interference conditions. This motivates the use of Q-learning for adaptive interference management in mmWave networks in this work. 

\subsection{Related Work}
Besides game-theoretic approaches, another line of research \cite{galindo2010distributed,van2012power,kar2013distributed
,ghadimi2017reinforcement, meng2019power, nasir2019multi, naderializadeh2021resource} has focused on learning-based methods. Galindo-Serrano {\it et al.} \cite{galindo2010distributed} considered the interference  control problem in cognitive radio (CR) systems where a set of CR UEs aim to maximize their own throughput while ensuring that the aggregate interference to the primary licensed UEs does not exceed a threshold. A decentralized Q-learning algorithm was proposed based on partial observation of the interference state. In \cite{kar2013distributed}, Kar {\it et al.} presented and analyzed a distributed reinforcement learning (RL) algorithm for collaborative multi-agent Markov decision processes. 
In particular, a distributed variant
of Q-learning was proposed where
each agent sequentially refines its learning parameters  based on the local payoff data and the information
received from neighboring agents.  Despite these classical methods, the use of deep neural networks (DNNs) as function approximators has also gained tremendous attention recently\cite{ghadimi2017reinforcement, meng2019power, nasir2019multi, naderializadeh2021resource}. 
Several works \cite{ghadimi2017reinforcement, nasir2019multi, naderializadeh2021resource} are relevant along this direction. 
Ghadimi {\it et al.}  \cite{ghadimi2017reinforcement} proposed a deep RL approach for downlink power control and rate adaptation in order to optimize a  utility function based on partial observability of the system state where each user is given an equal share of the bandwidth.  In~\cite{nasir2019multi}, a scalable and distributed multi-agent deep RL framework was proposed for transmit power control in cellular networks by assuming that the transmitters can obtain the CSI and QoS information from neighboring transmitters
Most relevant to our work is~\cite{naderializadeh2021resource} where a deep RL-based distributed power allocation and UE scheduling algorithm was proposed using a centralized training and distributed execution paradigm. Each access point can exchange instantaneous interference measurement with its neighbors. 

There are, however, several drawbacks of the aforementioned works. First, most of these works assume data exchange among BSs in the vicinity. This may not be practical if these BSs belong to different operators. How these approaches perform in such a fully distributed scenario also needs to be evaluated. Second, the performance of the deep RL-based approaches depend heavily  on the size and quality of the training dataset. However, it is not straightforward how these datasets can be obtained in advance. Offline training of the DNNs usually takes a significant amount of time which may not be suitable for delay-critical wireless systems. Third, models trained for a specific network configuration, e.g., number of BS/UEs, topology, fading etc., may not generalize well to other configurations, thus presenting scalability and robustness issues for these approaches.


In this paper, we present a general framework for distributed payoff optimization in non-cooperative mmWave networks and propose a Q-learning-based beam scheduling and power allocation approach using an independent modeling for each agent (i.e., BS) with a simple tabular representation of action-state values.
The proposed approach has lower complexity and better scalability than most deep RL-based approaches and is robust to network configuration change. At a similar complexity level, the proposed approach outperforms the game-based approach~\cite{zhang2020non} devised for the same problem.  Experiment results demonstrate that the proposed approach achieves a similar performance in the high SINR regime to the game-based approach and beats the game-based approach by a large margin in the relatively low SINR regime.


\section{Problem Formulation}

\subsection{System Description} 
\begin{figure}[t]
\centering
\includegraphics[width=0.4\textwidth]{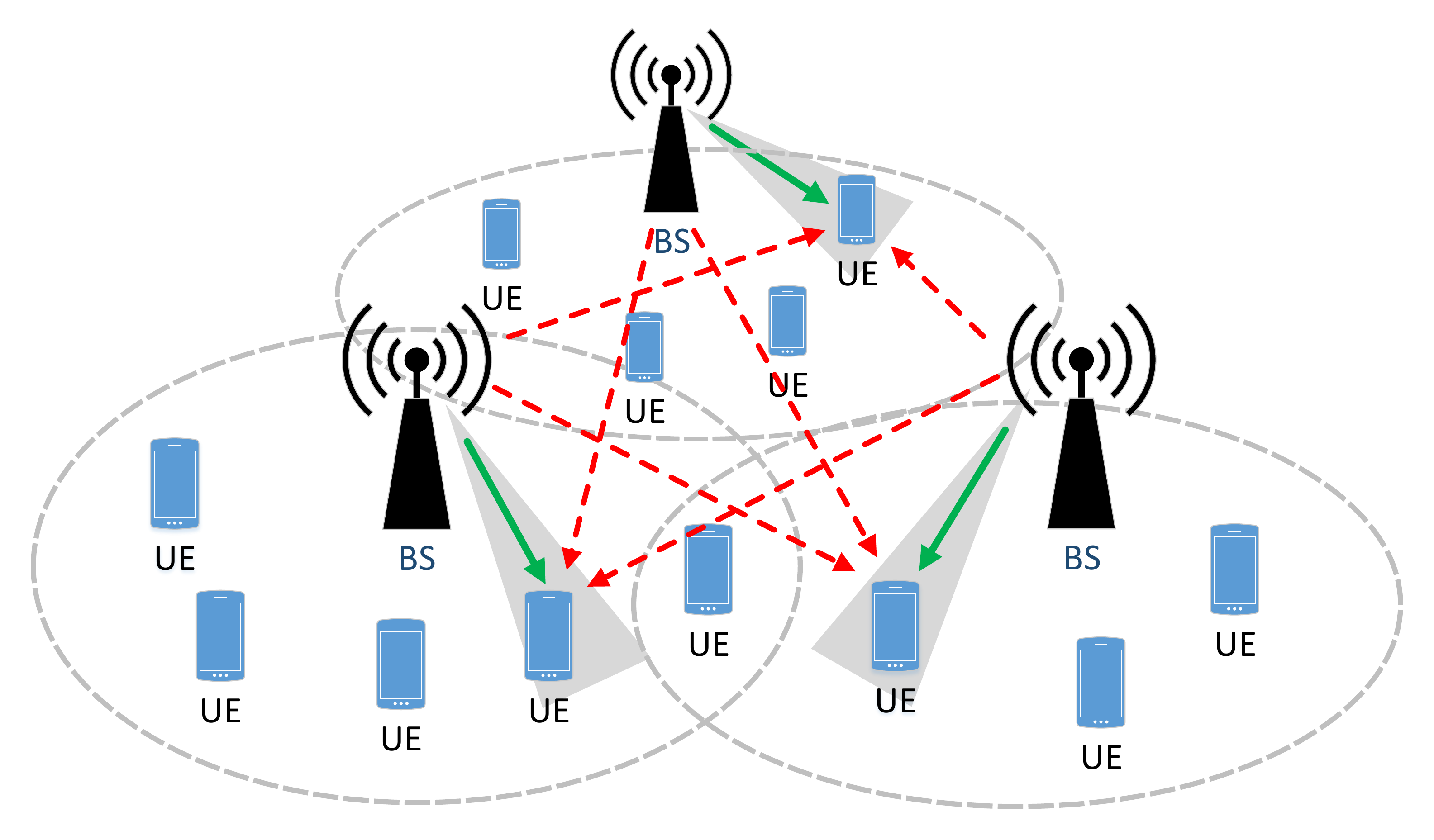}
\vspace{-0.1cm}
\caption{\small A cellular network consisting of $M$ BSs and $K$ UEs where each BS is associated with four UEs. The solid green lines represent the data links and the dashed red lines represent the interfering links.}
\label{fig: system}
\end{figure}

\begin{figure}[ht]
\centering
\includegraphics[width=0.44\textwidth]{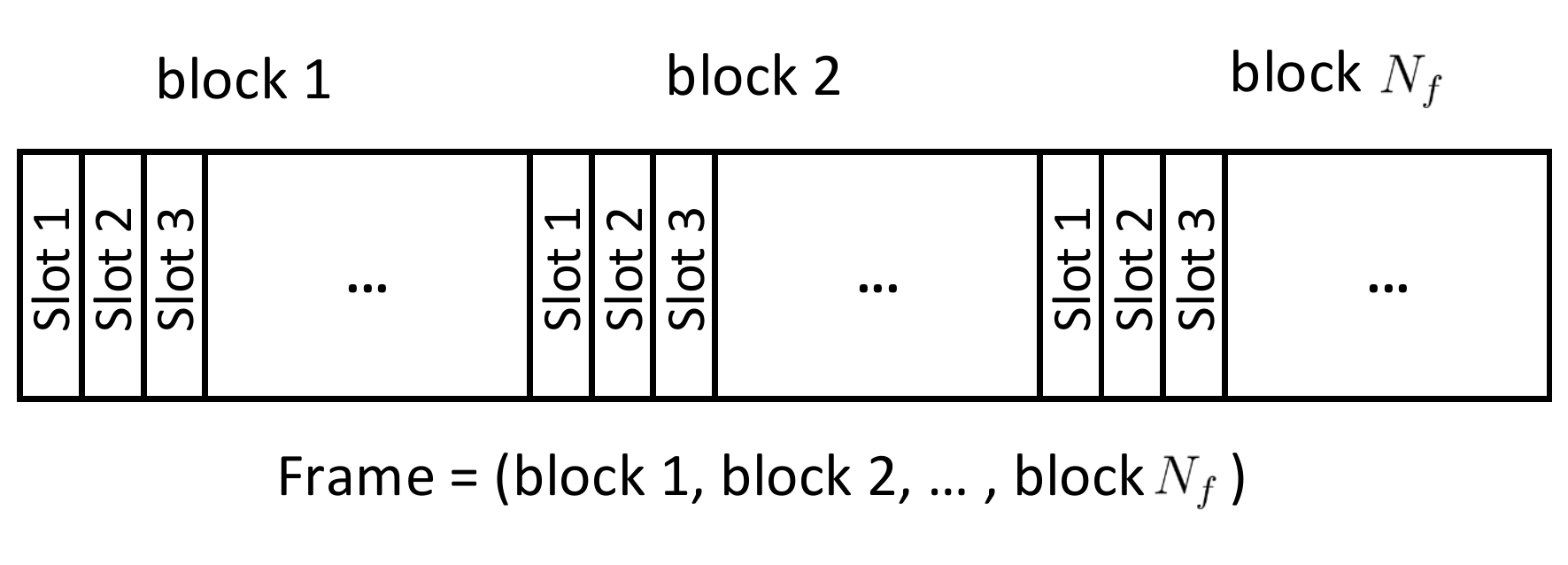}
\vspace{-0.3cm}
\caption{\small Frame structure. Each frame contains  $N_f$ blocks and each block contains $N_{b}$ slots.}
\label{fig: frame}
\end{figure}

Consider a cellular network with $M$ BSs $\Mc \eqdef \{1,2,...,M\}$ and $K$ UEs $\Kc \eqdef \{1,2,...,K\}$ as shown in Fig.~\ref{fig: system}. 
Each BS belongs to a different service  operator and is responsible for serving a set of $|\Kc_i|=K_i$ UEs within its coverage area.  We assume that each UE is served by exactly one BS and each BS can serve at most one UE at any given time. This means that $ \Kc_i \ne \emptyset, \forall i\in\Mc$, $\Kc_i\cap \Kc_{j}=\emptyset,\forall i\ne j$, and $ \cup_{i\in \Mc}\Kc_i=\Kc$. 
The BS-UE association is assumed to be determined by some exogenous mechanism and is fixed during the considered scheduling process. The system operates synchronously over a shared unlicensed spectrum of bandwidth $W$ Hz with a center frequency at $W_c$ Hz.
We use a frame structure as shown in Fig.~\ref{fig: frame}. Each time \emph{frame} contains $N_f$ \emph{blocks} and each block contains $N_b$ time \emph{slots} where each slot has a  duration of $T_s$ seconds. Therefore, each frame has a duration $T_f = N_fN_bT_s$ seconds and each block has duration $T_b = N_bT_s$ seconds. 
Beam and UE scheduling happens in each block of the frame which means that the beam and UE selection will stay unchanged during each block but will possibly change over different blocks. The BSs and UEs are equipped with directional antennas which are characterized by a  keyhole antenna model that is commonly used in the literature (e.g., by \cite{kiese2009connectivity,ramanathan2001performance}). The keyhole model has a constant main-lobe radiation gain $G^{\rm max}$ and a constant side-lobe gain $G^{\rm min}$. 
In particular, the antenna gain $G(\theta)$ in the direction $\theta$ is 
\begin{align}
G(\theta) = \left\{\begin{array}{cc} G^{\rm max}, & |\theta| \leq  \Theta/2 \vspace{0.1cm} \\
G^{\rm min}, & |\theta| > {\Theta}/{2} 
\end{array} \right.
\end{align}
where $\Theta$ is the beamwidth. The antenna also has a total radiation gain of $E$, i.e., $\Theta G^{\rm max }+(360^{\circ}-\Theta)G^{\rm min}=E.$
We further use $G^{\rm BS}_{j,i}$ and $G^{\rm UE}_{j,i}$ to respectively represent the antenna gain of $\sfb_i$ and $\sfu_j$ along the direction connecting $\sfb_i$ and $\sfu_j$. The \emph{main to side-lobe gain ratio (MSR)} is defined as
$\textrm{MSR} \eqdef 10 \lg \left( {G^{\rm max}}/{G^{\rm min}}\right)$.
A large MSR means that the antenna has strong radiation in the main-lobe while a small MSR implies energy leakage in the side-lobe. 
Due to the proximity of locations, the BSs may interfere with the UEs associated with other BSs. For $\sfb_i$, let $\sfu_{j_i}(j_i\in \Kc_i)$ be the UE selected by $\sfb_i$ to transmit to. Also, for any $\sfu_j$, let $\sfb_{i_j}$ be the BS that $\sfu_j $ is associated with $(j\in \Kc_{i_j})$. 
The Signal-to-Interference-Noise-Ratio (SINR) at $\sfu_j$ can be written as
\be
\label{eq: sinr at UE j} 
\textrm{SINR}_{j,i_j}=\frac{p_{j,i_j}G_{j,i_j}^{\rm UE}G_{j,i_j}^{\rm BS}|h_{j,i_j}|^2d_{j,i_j}^{-\eta}   }{\sum_{\ell\in\Mc, \ell\ne i  }p_{j_{\ell},\ell}G_{j,\ell}^{\rm UE}G_{j,\ell}^{\rm BS}|h_{j,\ell}|^2d_{j,\ell}^{-\eta}    +\sigma^2},
\ee
where $p_{j,i}$ denotes the transmit power of $\sfb_i$ to $\sfu_j$ if $\sfu_j$ is served by $\sfb_i$; $\eta$ is the path-loss factor; $\sigma^2=N_0W$ is the power of the random Gaussian noise ($N_0$ is the noise power spectrum density); $h_{j,i}$ is the small-scale fading between $\sfu_j$ and $\sfb_i$ which is assumed to follow the Nakagami-m distribution\cite{1318888} with probability density 
\be
\label{eq: Nakagam-m}
 f(h;\mu,\Omega)=\frac{2\mu^{\mu}}{\Gamma(\mu)\Omega^{\mu}}h^{2\mu-1}\textrm{exp}\left(-\frac{\mu}{\Omega}h^2\right),\;h\ge 0,
\ee
where $\mu \eqdef \frac{\mathbb{E}[h^2]^2}{\textrm{Var}(h^2)}$, $\Omega \eqdef\mathbb{E}[h^2]$ and $\Gamma(\cdot)$ is the Gamma function. We assume a block fading channel where the  fading coefficients stay unchanged during each frame and are i.i.d. over different frames\footnote{We do not consider UE mobility in this paper. However, the proposed approach applies to the case when UEs may move  slowly such that the channel gains do not change violently over different frames.}. We further define the \emph{equivalent channel gain} $g_{j,i_j}$ between $\sfu_j $ and $\sfb_{i_j}$ as $g_{j,i_j} \eqdef \textrm{SINR}_{j,i_j}/p_{j,i_j}$ if $\sfu_j$ is scheduled and $p_{j,i_j}> 0$.

\subsection{Payoff Maximization}
\label{subsection: Payoff Maximization}
Each BS is subject to an instantaneous peak transmit (TX) power constraint in each slot, i.e.,  $\sum_{j\in \Kc_i} p_{j,i} \le  p_i^{\rm max}$. Since it is assumed that at most one UE can be scheduled at a time, we have $p_{j_i,i}\le p_i^{\rm max}$ where $\sfu_{j_i}$ is the scheduled UE by $\sfb_i$. Let $\bm{p}\eqdef \{p_{j_i,i}\}_{i\in \Mc}$ denote the TX powers of the BSs to their respective scheduled UEs. We consider a general form of {\it payoff} function (for a unit time duration of one second) for each BS which is defined as
\be 
\label{eq: BS utility}
R_i(\bm{p}) \eqdef \alpha_i W\log\left( 1+ \textrm{SINR}_{j_i,i}\right) - \beta_ip_{j_i,i},
\ee 
i.e., the payoff of $\sfb_i$ is the throughput of its scheduled UE (weighted by $\alpha_i$) plus a power penalizing  term (weighted by $\beta_i$). The weights $\alpha_i,\beta_i\ge 0$ can be tuned manually or determined 
using some algorithms\footnote{An exmaple is presented in Section \ref{subsection: App to Lyapunov OPT} where the weights are determined by the queue values derived from the Lyapunov optimization framework.} in order to find a desirable trade-off between throughput and power consumption. In particular, the ratio $\alpha_i/\beta_i$ determines the relative importance of throughput maximization to power consumption. If $\alpha_i/\beta_i$ is very large, eq. (\ref{eq: BS utility}) becomes equivalent to maximizing the throughput $R_i(\bm{p})\approx \alpha_i W\log\left( 1+ \textrm{SINR}_{j_i,i}\right)$. Note that the solution becomes trivial when either $\alpha_i$ or $\beta_i$ is equal to zero. 
For any given set of scheduled UEs $\{j_i\}_{i\in \Mc}$,  we aim to find efficient power allocation schemes to maximize the sum payoff $R(\bm{p})$ of all BSs $
R\left(\bm{p}\right)\eqdef \sum_{i\in \Mc}R_i(\bm{p}).$ 
Let $\bm{p}(t)$ be the power allocation profile in slot $t$. Then our goal is to maximize the long-term average payoff 
\be 
\label{eq: avg sum payoff}
\bar{R} =\lim_{T\to \infty} \frac{1}{T}\sum_{t=1}^T R\left(\bm{p}(t)\right).
\ee
The challenge lies in that this sum payoff maximization problem must be solved in a distributed manner, that is, there is no centralized control or coordination among the BSs as they belong to different service operators. It should also be noted that the above formulation is not particular to any specific scheduling method so new scheduling methods can be developed under the same framework and be effectively evaluated by comparing to previous methods.

\section{Proposed Approach}
Under the general formulation in Section~\ref{subsection: Payoff Maximization}, we propose to solve the payoff maximization problem (\ref{eq: avg sum payoff}) using Q-learning by modeling each BS as an independent learning agent that interacts with the radio environment which is governed by the collective behavior of all agents and  channel uncertainty. By properly defining the state space and rewards, the proposed  learning-based beam scheduling and power allocation is shown to be able to outperform the game-theoretic (GT) approach~\cite{zhang2020non} -- a previously developed iterative power allocation algorithm for the considered mmWave scheduling problem, especially in the interference-limited regime. In  
the following, we first present a brief background of Q-learning and then proceed to the description of the proposed approach.

\subsection{Q-learning Preliminary}
In RL, an agent interacts with the environment by making decisions that may affect the state of the environment in a sequence of discrete time steps. In particular,  at time $t$, based on the observation of the current state $s^{(t)}$ of the environment, the agent takes an \emph{action} $a^{(t)}$ according to a {\it policy} $\pi$ as $a^{(t)} \sim \pi(\cdot|s^{(t)})$ with a special case of being deterministic with $a^{(t)} =\pi(s^{(t)})$. 
After taking the action, the agent  receives an immediate reward $r^{(t)}$ which indicates the quality of the chosen action $a^{(t)}$ in state $s^{(t)}$. As a result of the above interaction, the environment transitions to a new state $s^{(t+1)}$. 
The goal of RL is to  maximize the agent's  long-term expected reward $G^{(t)}$ defined as $G^{(t)}\eqdef \sum_{k=0}^{\infty}\gamma^k r^{(t+k+1)}$, where $\gamma$ is the discount factor which indicates the importance of future rewards.   
Model-free RL aims to find a an optimal policy $\pi^*$ that maximizes the expected reward $G^{(t)}$ by learning  directly from the agent-environment interactions represented by a set of quadruples $\left\{(s^{(\ell)},a^{(\ell)},r^{(\ell)},s^{(\ell+1)}): \ell\le t \right\}$ 
called \emph{experience} (up to time $t$), without any specific knowledge of the underlying transition probabilities of the environment. 

Q-learning is a model-free off-policy learning algorithm for estimating the optimal action-state values $q_{*}(a,s)$ for each action-state pair $(a,s)\in \Ac\times \Sc$ ($\Ac$ and $\Sc$ denote the action and state space respectively).  Let $Q(s,a)$ denote an estimate of $q_{*}(a,s)$. At time $t$, the agent chooses its action using the $\epsilon$-greedy action selection method, that is, with a small probability $\epsilon$ (also termed as exploration rate), the agent chooses a random action  in $\Ac$; else it chooses a greedy action $a^{(t)}=\arg\max_{a\in \Ac}Q(a,s^{(t)})$. After the selection, the action-state values are updated according to 
\begin{align}
\label{eq: QL update 1}
&Q\left(a^{(t)},s^{(t)}\right)\leftarrow  (1-l_r)Q\left(a^{(t)},s^{(t)}\right)\notag \\
&\qquad \qquad  \qquad \quad   \;  + l_r \left( r^{(t)} + \gamma \max_{a\in \mathcal{A}}Q\left(a, s^{(t+1)}\right)\right),
\end{align}
and $Q\left(a,s\right)$ does not update if $(a,s)\ne (a^{(t)},s^{(t)})$. $l_r\in (0,1]$ is the learning rate which determines to what extent the new estimate $r^{(t)} + \gamma \max_{a\in \mathcal{A}}Q\left(s^{(t+1)},a\right)$ overrides the old estimate $Q\left(a^{(t)},s^{(t)}\right)$. 
Q-learning usually employs a tabular representation $[Q(a,s)]_{|\Ac|\times |\Sc|}$, the Q-table, to store  the estimated action-state values. For continuous action or state spaces, neural networks can be used to approximate the action-state values\cite{santamaria1997experiments,gross1998neural}. 
For a stationary underlying transition model, the Q-learning  algorithm  converges to the optimal policy with probability one asymptotically if the learning rate $l_r(t)$ at time $t$ satisfies $\sum_{t=1}^{\infty}l_r(t)= \infty, \sum_{t=1}^{\infty}l_r(t)^2 <\infty$. For optimizing an expected reward over a finite horizon $T$, a constant learning rate $l_r$ can be used.

\subsection{Why Q-learning?}
\label{subsection: why using QL}
One key feature of the learning-based methods, specifically  Q-learning 
that will be using in this work,  is the ability to adapt by learning from experience and exploring, going beyond the mere greedy nature of the game-based methods. One major challenge in the considered mmWave scheduling problem is how to handle the strong interference due to the lack of centralized coordination of beams. Being purely greedy in this scenario can potentially hurt the overall performance. In particular, if we model each BS as a non-cooperative game player that myopically focuses on maximizing its own payoff (say the throughput) in each slot, then each BS will always choose the maximum power to transmit since it gets maximum throughput from this decision. However, if the beams of different BSs overlap, there will be very strong interference at the scheduled UEs, which in turn yields a small network-level payoff (also see Section~\ref{Baseline Scheme} for a detailed analysis). What is even worse is that this situation can happen over and over again as the BSs do not learn from these bad experience. In contrast, if we model each BS as an Q-learning agent, the case of overlapping beams can still occur. However, the decisions of the BSs can be very different from the game-based methods. First, each BS can explore non-greedy actions using the $\epsilon$-greedy action selection, partly avoiding the maximum TX power dilemma. Second, each BS can also learn from its past experience to improve the performance. If the overlapping beam situation happens and the BS has chosen the maximum power, then it will receive a small reward ($r^{(t)}$ in (\ref{eq: QL update 1})) due to strong inter-cell interference. This will inform the BS to avoid using maximum power in similar situations in the future and thus improves the long-term throughput performance.

\subsection{Proposed Beam Scheduling and Power Allocation}
\label{subsection: proposed approach}
Due to the adaptation ability of  Q-learning as described in the previous section and its simplicity, we focus on applying the classical Q-learning algorithm to the considered mmWave scheduling problem. In particular, we model each non-cooperative BS as an independent learning agent that implements the Q-learning algorithm presented in the previous section in parallel. The key Q-learning components for each agent are defined as follows.
\begin{itemize}
    \item[{\it 1)}]{{\it Environment}:} Each agent interacts with the physical radio environment governed by the collective behaviors, e.g., UE scheduling, TX powers, beam generation, etc., of the BSs subject to random channel realization.
    
    \item[{\it 2)}]{{\it Action}:} The action for $\sfb_i$ in each slot is the TX power $p_{j_i,i}^{(t)}$. To use the tabular representation of Q-learning, the action and state spaces must be discrete. Therefore, we quantize the TX power range  $[0,p_i^{\rm max}]$ uniformly into $P_q$ discrete levels $\Pc_q=\{ p_i^1,p_i^2,\cdots,p_i^{P_q}\}$ to represent the action space where 
    \be p_i^{j}=(j-1)\frac{p_i^{\rm max}}{P_q-1},\;  j\in\{1,\cdots,P_q\}.\ee 
    This means $p_i^1=0$ and $p_i^{P_q}=p_i^{\rm max}$. The same uniform power quantization is used by all BSs.
    
    \item[{\it 3)}]{{\it Observation}:} Each BS's observation of the environment is defined as the received (RX) interference (plus noise) at its scheduled UE. Let $I_{j_i,i}$ denote the RX interference at $\sfu_{j_i}$. Suppose $I_{j_i,i}$ follows a (possibly unknown) distribution $\Dc_{j_i,i}$ over the range $[I_{j_i,i}^{\rm min},I_{j_i,i}^{\rm max} ]$ with $I_{j_i,i}^{\rm min}$ and $I_{j_i,i}^{\rm max}$ being the minimum and maximum possible interference respectively. The RX interference also needs to be quantized in order 
    to be represented by a discrete {\it state}. We propose a {\it percentile-based} quantization method as follows. We first derive $I_q$ percentiles $\Ic_q=\{I_1,I_2,\cdots, I_{I_q} \}$ over the distribution $\Dc_{j_i,i}$. This means that the probability that $I_{j_i,i}$ falls into any interval $(I_j,I_{j+1}]$ is identical and is equal to $1/I_q,\forall j\in\{1,\cdots,I_q-1\}$. 
    If the measured interference $I_{j_i,i}$ fall into the interval $(I_j,I_{j+1}]$, we say that the observation of $\sfb_i$ is `state $j$'. Therefore, the state space of $\sfb_i$ can be represented by $\Sc_i=\{1,2,\cdots,I_q\}$. The proposed quantization method guarantees that each state will be visited approximately the same number of times in the long run. An illustration of the percentile-based quantization method with $I_q=10$ states is shown in Fig.~\ref{fig: interf. hist}. We assume all BSs use the same number of states. It should be noted that the UE interference distributions are not know by the BSs so they have to be estimated, after which the above state quantization can be conducted. 
    \begin{figure}[ht]
    \centering
    \includegraphics[width=0.39\textwidth]{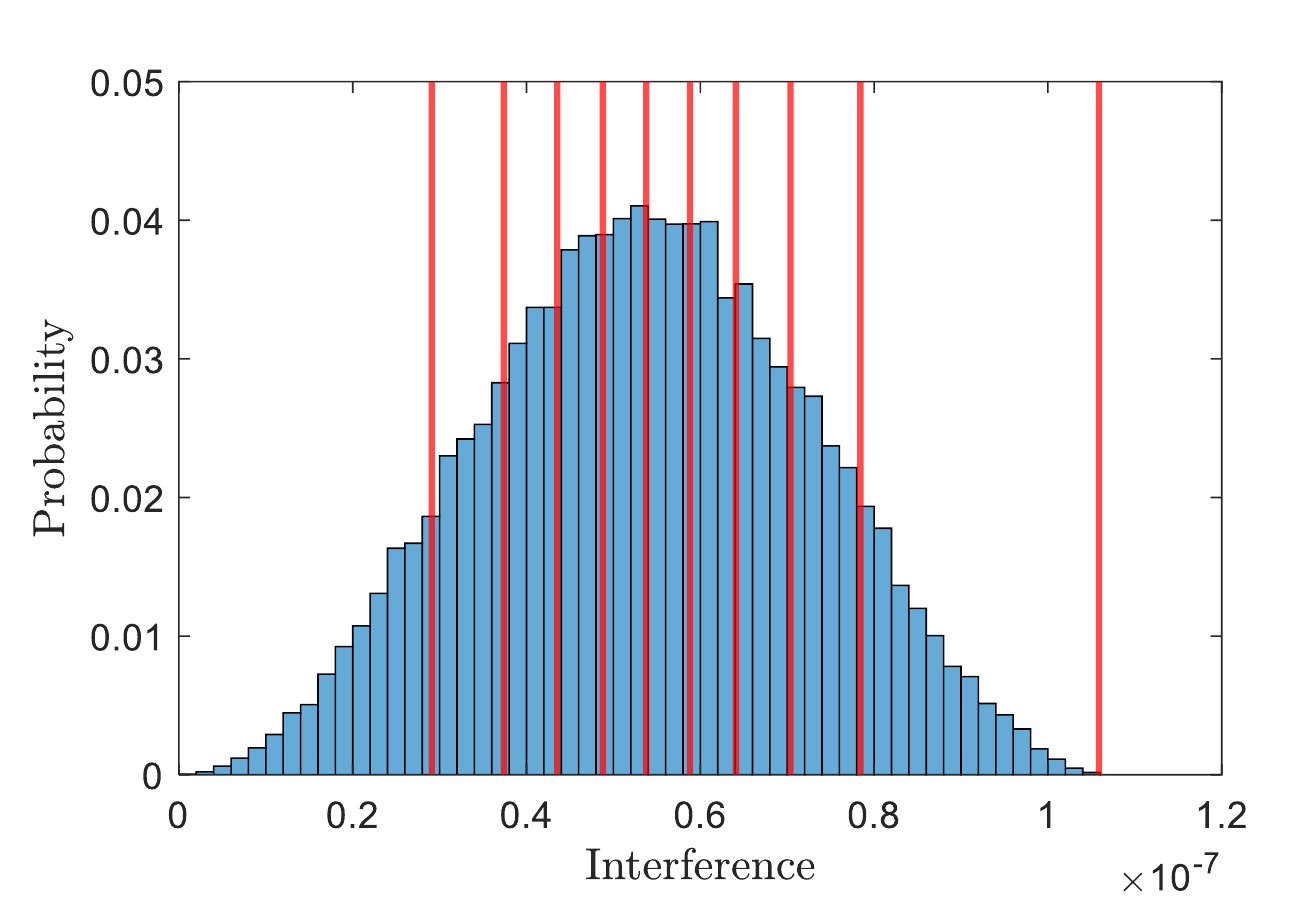}
    \vspace{-0.3cm}
    \caption{\small Percentile-based interference quantization with ten levels based on an empirical interference distribution. }
    \label{fig: interf. hist} 
    \end{figure}
    \item[{\it 4)}]{{\it Reward}:} The reward of $\sfb_i$ in slot $t$ is defined as
    \be 
    \label{eq: BS i reward each slot}
    r_i^{(t)} \eqdef \alpha_i\left(T_s W\log\big(1+\textrm{SINR}_{j_i,i}^{(t)}\big)\right)-\beta_i \left(T_sp_{j_i,i}^{(t)}\right),
    \ee 
    where $\textrm{SINR}_{j_i,i}^{(t)}$ is the SINR at $\sfu_{j_i}$ in slot $t$. The goal of $\sfb_i$ is to maximize the long-term expected (discounted) reward 
    \be 
    \label{eq: BS i QL expected greward}
    G_i^{(t)} = \sum_{k=0}^{\infty}\gamma^{k}r_i^{(t+k+1)}
    \ee 
    starting from any time $t$. It should be noted that when the discount factor $\gamma$ is close to 1, eq. (\ref{eq: BS i QL expected greward}) 
    can be used to approximate (\ref{eq: avg sum payoff}) after averaging over time. 
\end{itemize}

With the above definitions of the action, observation/state and the reward function, we propose to solve the sum payoff maximization problem (\ref{eq: avg sum payoff}) by letting each BS `selfishly' maximize its own average payoff $\bar{R}_i\eqdef \lim_{T\to \infty}\frac{1}{T}R_i(\bm{p}(t))$. To do this, we model each BS as an independent learning agent implementing the $\epsilon$-greedy action selection method with the goal of optimizing its long-term expected reward (\ref{eq: BS i QL expected greward}). For a any finite $T$ and $\gamma\approx 1$, optimizing $\bar{R}_i=\frac{1}{T}\sum_{t=1}^T R_i(\bm{p}(t))$ becomes equivalent to optimizing (\ref{eq: BS i QL expected greward}). Therefore, we have provided a  fully distributed approach 
using Q-learning in a multi-agent scenario. The proposed beam scheduling and power allocation scheme consists of  a training phase followed by an execution phase, which are described as follows. 

\textbf{Training Phase:}  
This phase is to estimate the empirical distribution of the RX interference at each UE so that the interference quantization can be done during the scheduling execution phase. 
In particular, for the set of scheduled UEs $\{j_i\}_{i\in \Mc}$, we run $T_{\rm train}$ frames of `simulated scheduling' in which the TX powers of the BSs are chosen randomly from $\Pc_q$ in each slot  
and the wireless channels are subject to change from frame to frame. 
We record the interference at each scheduled $\sfu_{j_i}$ in all the training frames and derive an empirical interference distribution $\hat{\Dc}_{j_i,i}$, which will be used to quantize the RX interference in the execution phase. Note that during the training phase, although the powers are randomly selected, the BS/UEs still achieve some data throughput in each slot. Moreover, this training phase  only needs to be done once before the `real' scheduling begins, so the overhead induced by this phase becomes negligible if we consider the scheduling problem over a large number of frames.

\textbf{Execution Phase:} Beam scheduling and power allocation happen in this phase where the frame structure of Fig.~\ref{fig: frame} is used. Since we do not consider UE scheduling in this paper, the UEs can be scheduled randomly or in a round robin manner in different blocks. Therefore, we focus on the application of the proposed scheduling approach in one block. Each BS implements the Q-learning algorithm as follows. At the beginning of slot $t$, based on the current state which is defined as the quantized RX interference at $\sfu_{j_i}$ in slot $t-1$ (this interference is measured by $\sfu_{j_i}$ and then feedback to $\sfb_i$), $\sfb_i$ chooses TX power $p_{j_i,i}^{(t)}$ according to the $\epsilon$-greedy action selection method,  it then generates a beam towards $\sfu_{j_i}$ and starts the data transmission. Note that no beams will be generated if $p_{j_i,i}^{(t)}=0$. After the beam generation, $\sfb_i$ updates its Q-table according to 
(\ref{eq: QL update 1}) where the next state $s^{(t+1)}$ is defined as the quantized RX interference at $\sfu_{j_i}$ in slot $t$ (after the power selection), and the reward $r_i^{(t)}$ is defined 
in (\ref{eq: BS i reward each slot}). The above process is repeated until the end of the current block. The proposed approach, performed in one block, is summarized in Algorithm~\ref{algorithm:1}. We use the notation  $[n]\eqdef \{1,2,\cdots, n\}$ for $n\in \mathbb{N}$.
\begin{algorithm}
\setstretch{1.2}  
  \caption{Proposed Beam Scheduling \& Power Allocation: Execution Phase}
  \label{algorithm:1}
  \begin{algorithmic}[1]
  \item {\bf Input}: $P_q,I_q,N_b,\alpha,\beta,\gamma,\epsilon$ and $l_r$.
  \item {\bf Initialization}: Each $\sfb_i$ randomly picks $\sfu_{j_i}$ and initialize Q-table as $Q_{i}(a,s)=1,\forall (a,s)\in [P_q]\times [I_q]$. Set $t=1$.
  \item {\it Step 1:} $\sfb_i$ chooses TX power $p_{j_i,i}^{(t)}$ in slot $t$ according to
  \begin{align}
p_{j_i,i}^{(t)} = \left\{\begin{array}{ll} 
\textrm{randomly pick from } \Pc_q, & \textrm{w.p. } \epsilon \\
p_{\hat{a}}, \hat{a} = \arg\max\limits_{a\in[P_q] } Q_i\big(a,s^{(t)}\big)   , & \textrm{w.p. } 1-\epsilon\notag
\end{array} \right.
\end{align}
$\sfb_i$ generates a beam towards $\sfu_{j_i,i}$ if $p_{j_i,i}^{(t)}\ne 0$.
  \item {\it Step 2:} Each $\sfb_i$ updates its Q-table according to:  let $Q_i(a,s)\leftarrow Q_i(a,s)$, if $(a,s)\ne (a^{(t)},s^{(t)})$, and 
  let $Q_i(a,s)\leftarrow (1-l_r)Q_i(a,s) +  l_r\left(r_i^{(t)} + \gamma \max_{a\in [P_q]}Q_i\big(a, s^{(t+1)}\big)\right)$, if $(a,s) =(a^{(t)},s^{(t)})$.
  \item {\it Step 3:} $t\leftarrow t+1$. If $t\le N_b$, go back to {\it Step 1}, else stop.
  \item \textbf{Output}: Average reward of all BSs.
  \end{algorithmic}
\end{algorithm}

\begin{remark}
In Algorithm~\ref{algorithm:1},  the Q-tables of the BSs are initialized with 
all one matrices, i.e., the initial value estimate are set to $Q_i(a,s)=1,\forall a,s$. This is termed as the principle of {\it being optimistic in the face of uncertainty} which is widely used in value-based RL applications.
\end{remark}

\begin{remark}[Complexity]
For each BS, the storage complexity of the proposed algorithm is  $\Oc\left(\frac{KP_qI_q}{M}\right)$ (supposing each BS is associated with the same number of UEs) since each BS has to store a Q-table of size $P_q\times I_q$ for each of its $K/M$ associated UEs. In the execution phase, the implementation complexity per slot is $\Oc\left(\max\{P_q,I_q\}\right)$ which is due to the UE interference quantization ($\Oc(I_q)$) and greedy action selection ($\Oc(P_q)$). The Q-table update has complexity $\Oc(1)$. It can be seen that both the storage and implementation complexity scale linearly with the number of discrete powers and interference states, and the storage complexity also scales linearly   with the number of UEs. This linear scaling is acceptable in general. Our experiments in Section \ref{sec: experiment} show that the typical values of $P_q\approx 10,I_q\approx 20$ suffice to achieve the near-optimal (by letting $P_q,I_q$ being arbitrarily large) performance for the considered network in the experiment with four BSs and twelve UEs in total.
\end{remark}

\section{Experiment}
\label{sec: experiment}
\subsection{Experiment Setup}
\begin{figure}[ht]
\centering
\includegraphics[width=0.36\textwidth]{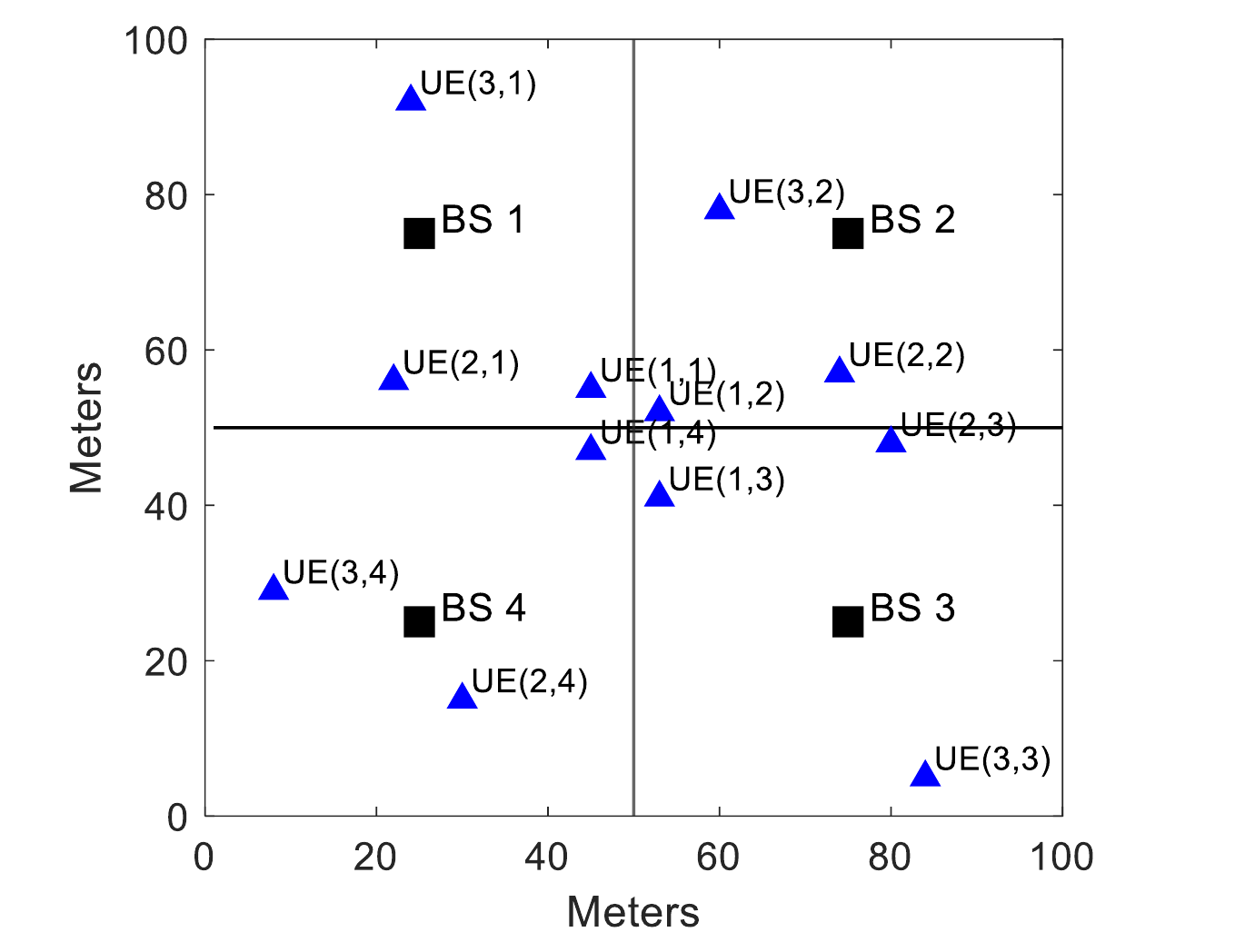}
\vspace{-0.2cm}
\caption{\small Locations of the BSs and UEs on a $100\times 100$ $\textrm{meter}^2$ planar grid. UE $(j,i)$ represents the $j^{\rm th}$ UE of $\sfb_i$.}
\label{fig: experiment position}
\end{figure}

Consider
a cellular network (
see Fig.~\ref{fig: experiment position}) with four BSs each belonging to different operator. Each BS is associated with three UEs located randomly in its coverage area. 
Let $l=20$ meters be the height of the BS antenna. UE antenna height is assumed to be zero. Therefore, the distance from $\sfb_i$ to $\sfu_j$ is equal to $d_{j,i} = \sqrt{ l^2+ \bar{d}_{j,i}^2}$ where $\bar{d}_{j,i}$ is the planar distance between $\sfb_i$ site and $\sfu_j$.
The system has a shared bandwidth of $W=400$ MHz with a center frequency $W_c=37$ GHz.  Each BS is subject to a peak instantaneous power constraint $p_i^{\rm max}=39$ dBm (7.94 Watt).  Noise power is calculated according to 
\be 
\sigma^2 \,(\textrm{dBm}) = 10\lg(\kappa_{ B}T_0\times 10^3) + \textrm{NR}\,\textrm{(dB)} + 10\lg W \notag
\ee
where $\kappa_{B}=1.38\times 10^{-23}$ J/K is  Boltzmann's constant, NR is the UE noise figure and $T_0$ is the temperature. Taking the typical values of $\textrm{NR}=1.5\, \textrm{dB}$ and $T_0 = 290\, \textrm{K}$, the total noise power over the 400  MHz bandwidth is equal to $\sigma^2 =-86.46 \,\textrm{dBm}$. We consider the beam scheduling and power allocation in one block with $N_b=100$ slots. Each slot has a duration of one milli-second. The physical environment and learning parameters are listed as follows:
\begin{table}[htbp]
\begin{center}
\begin{tabular}{|c"c|}
\hline
 \textrm{Parameter} & \textrm{Value}\\
 \thickhline 
 exploration rate $\epsilon$ & $0.05$\\
 \hline 
 discount factor $\gamma $ & $0.9$\\
 \hline 
learning rate $l_r$ & $0.1$\\
 \hline 
 $p_i^{\rm max},\forall i\in \Mc$ & 7.94 \textrm{Watt}\\
 \hline
  noise power $\sigma^2$ & $-86.46$ dBm \\
   \hline 
    pass loss $\eta$ & 4 \\
   \hline
    Nakagami fading $\Omega, \mu$ & $100,10^4$ \\
   \hline
 block size $N_b$ & $100$ slots\\
   \hline 
slot duration $T_s$ & 1 millisecond\\
\hline
BS antenna height $l$ & 20 meters\\
\hline
\end{tabular}
\label{tab: parameter list}
\end{center}
\end{table}

\subsection{Baseline Scheme}
\label{Baseline Scheme}
\textit{Game-Theoretic (GT) Power Allocation:}
In \cite{zhang2020non}, a non-cooperative game-based power allocation was proposed for distributed interference management in mmWave networks. In particular, each BS is treated as an independent {\it player} that selfishly attempts to maximize its own payoff, defined in the  form of (\ref{eq: BS utility}). A parallel power adaptation scheme was proposed based on the the concept of {\it best response}.  In each slot, $\sfb_i$ updates its power according to 
\be
  \label{eq: GT update}
   p_{j_i,i}^{(t+1)}= \left[\frac{\alpha_i W}{\beta_i}-\frac{1}{g_{j_i,i}^{(t)}}     \right]_0^{p_i^{\rm max}},
\ee  
where $g_{j_i,i}^{(t)}\eqdef  {G_{j_i,i}^{\rm BS} G_{j_i,i}^{\rm UE}|h_{j_i,i}|^2d_{j_i,i}^{-\eta}}/({I_{j_i,i}^{(t)} + \sigma^2}) $ is the equivalent channel gain between $\sfb_i$ and $\sfu_{j_i}$ in slot $t$. $g_{j_i,i}^{(t)}$ can be obtained by $\sfb_i$ by letting $\sfu_{j_i}$ measuring the RX interference (plus noise) $I_{j_i,i}^{(t)}+\sigma^2$ and then sending back to $\sfb_i$. The Euclidean projection operator $[\cdot]_a^b$ is defined as $[x]_a^b= a$ if $x<a$, $[x]_a^b= b$ if $x>b$ and $[x]_a^b= x$ if $x\in [a,b]$. The above power adaptation is proved to converge to Nash equilibrium under certain conditions.

\begin{figure}[ht]
\centering
\includegraphics[width=0.48\textwidth]{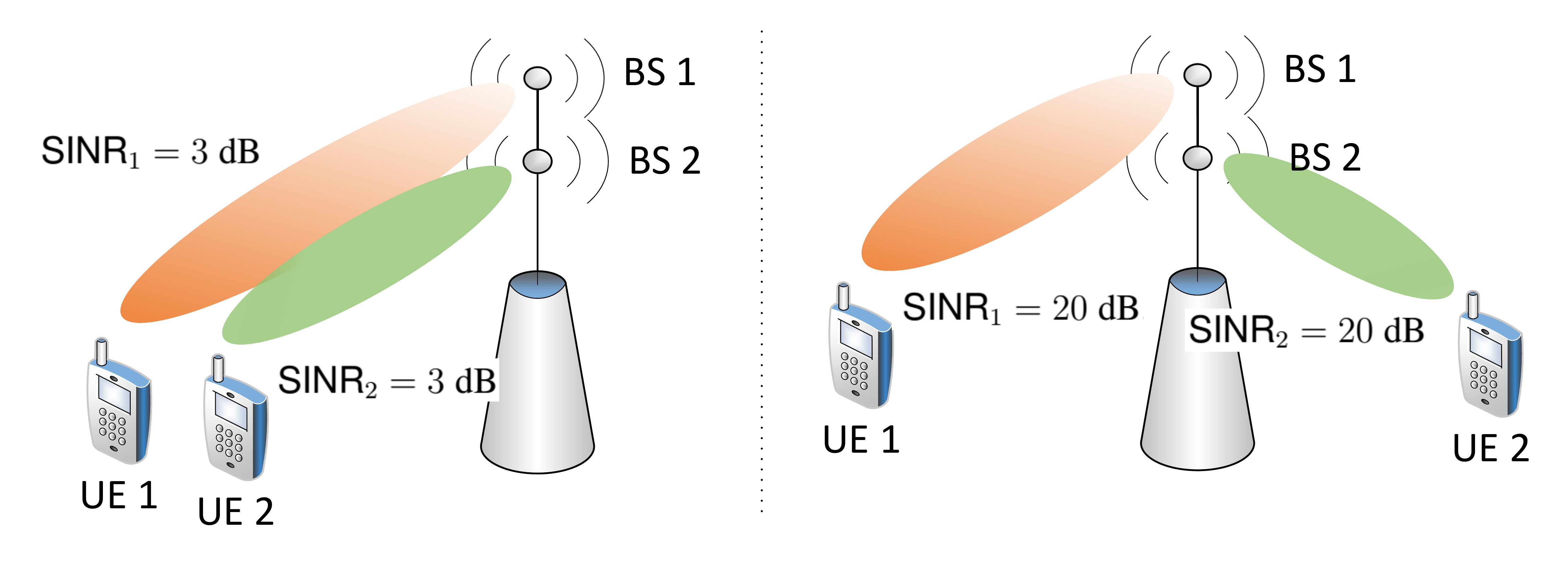}
\vspace{-0.2cm}
\caption{\small BS colocation and and closely located UEs (left).  There is strong interference due to beam overlapping. GT cannot distinguish the two cases.}
\label{fig: overlapping BS}
\end{figure}

{\it Drawback of the GT power allocation:} 
The GT power allocation has a disadvantage due to its greedy nature: it may perform poorly in the high interference regime. This is because, for example, for the case of $\beta_i \approx 0$, each BS only aims to maximize its own throughput. The solution to GT is always choosing the maximum power to transmit, regardless of the interference. This may cause severe interference if the scheduled UEs are close to each other or there is beam overlapping (See Fig.~\ref{fig: overlapping BS}),
and thus dampening the overall performance. However, our proposed Q-learning-based approach 
can overcome the above disadvantage by adapting to the physical environment (via observation and action-state value update) which is governed by the joint behaviors of all the agents. Each BS may make decisions other than maximum power based on the current interference state and its experience. 
For example, for the overlapping beam case, if all BSs are transmitting with high powers, being greedy by choosing a large TX power will emit a small reward as all UE are experiencing strong interference. By learning from the small reward, the Q-learning-based approach can shift to lower power to explore new possibilities of higher reward. However, the GT allocation will be greedy and unable to adapt. Another drawback of the GT  method is that it operates with continuous power which is infeasible in practice. However, quantization of TX power will inevitably  incur performance loss by the adaptation rule of (\ref{eq: GT update}). In section \ref{subsec: Experiment result}, we verify  the effect of multiple factors that affect the  performance of the proposed  approach and show that the performance can be significantly enhanced over GT.

\subsection{Experiment Result}
\label{subsec: Experiment result}
We compare the proposed approach with the GT power allocation and verify the effect of the reward weights $\alpha,\beta$, the number of power levels $P_q$ and interference states $I_q$ and the BS/UE antenna gain and beamwidth.  Throughout the experiment, we assume that all UEs have omnidirectional antennas.\footnote{Since varying the UE antenna MSR and beamwidth has a similar effect to that of the BS antenna, we use omnidirectional UEs in the experiment.} We fix $\alpha=1$ for all BSs and and let $\beta=0$ and $\beta=0.1W=4\times 10^7$ to verify its effect.
\subsubsection{Effect of $P_q$ and $I_q$}
\begin{figure*}[t]
    \centering
    \begin{subfigure}[h]{0.4\textwidth}
        \includegraphics[width=\textwidth]{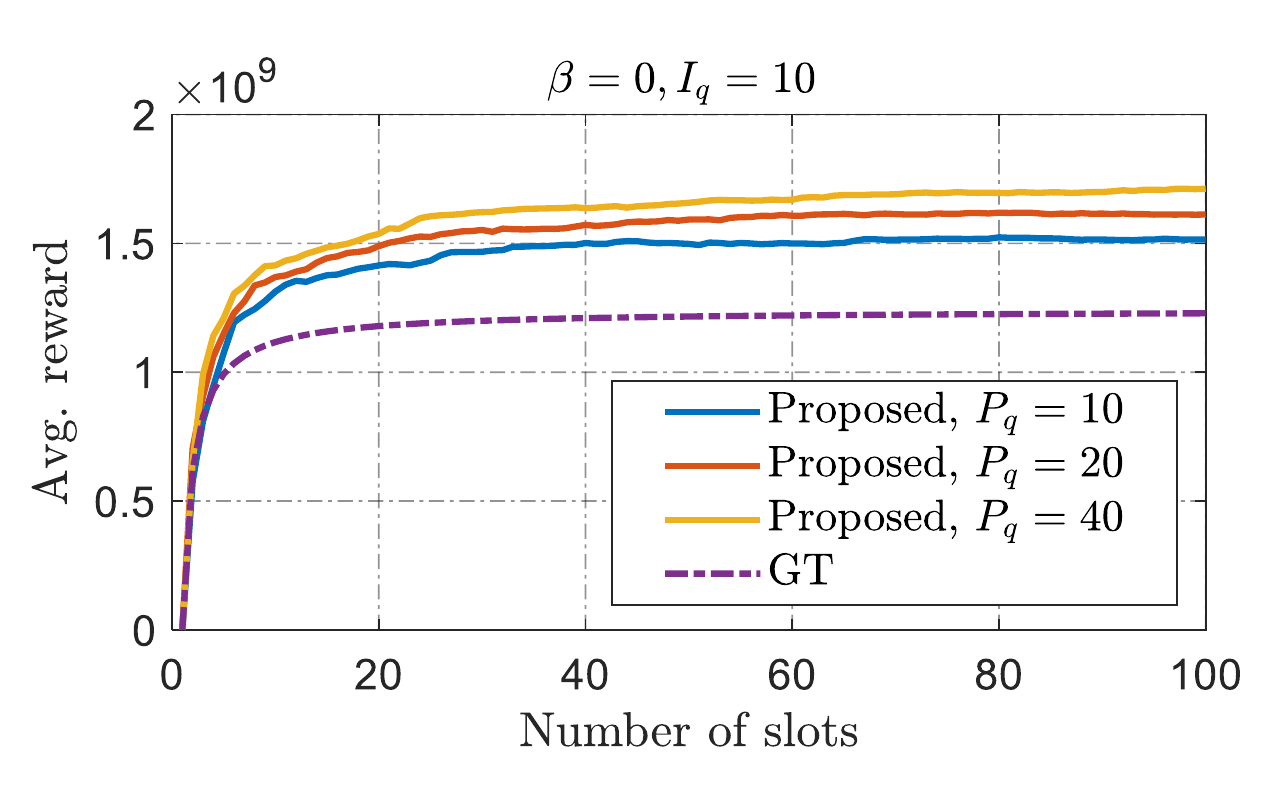}
        \vspace{-0.8cm}
        \caption{\small Effect of $P_q$ when $\beta =0$, $I_q=10$.}
        \label{subfig 11}
    \end{subfigure}
    \begin{subfigure}[h]{0.4\textwidth}
        \includegraphics[width=\textwidth]{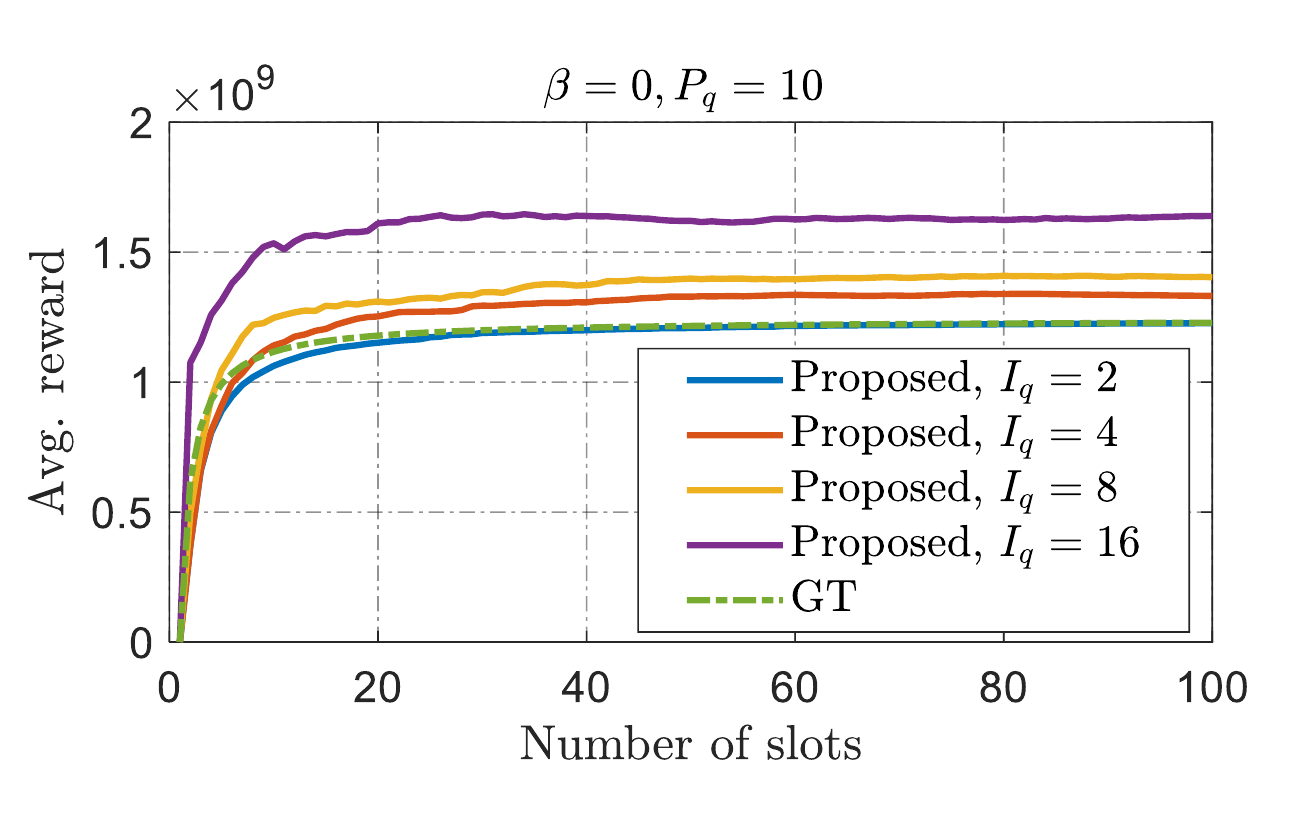}
        \vspace{-0.8cm}
        \caption{\small Effect of $I_q$ when $\beta =0$, $P_q=10$.}
        \label{subfig 12}
    \end{subfigure}
    \begin{subfigure}[h]{0.4\textwidth}
        \includegraphics[width=\textwidth]{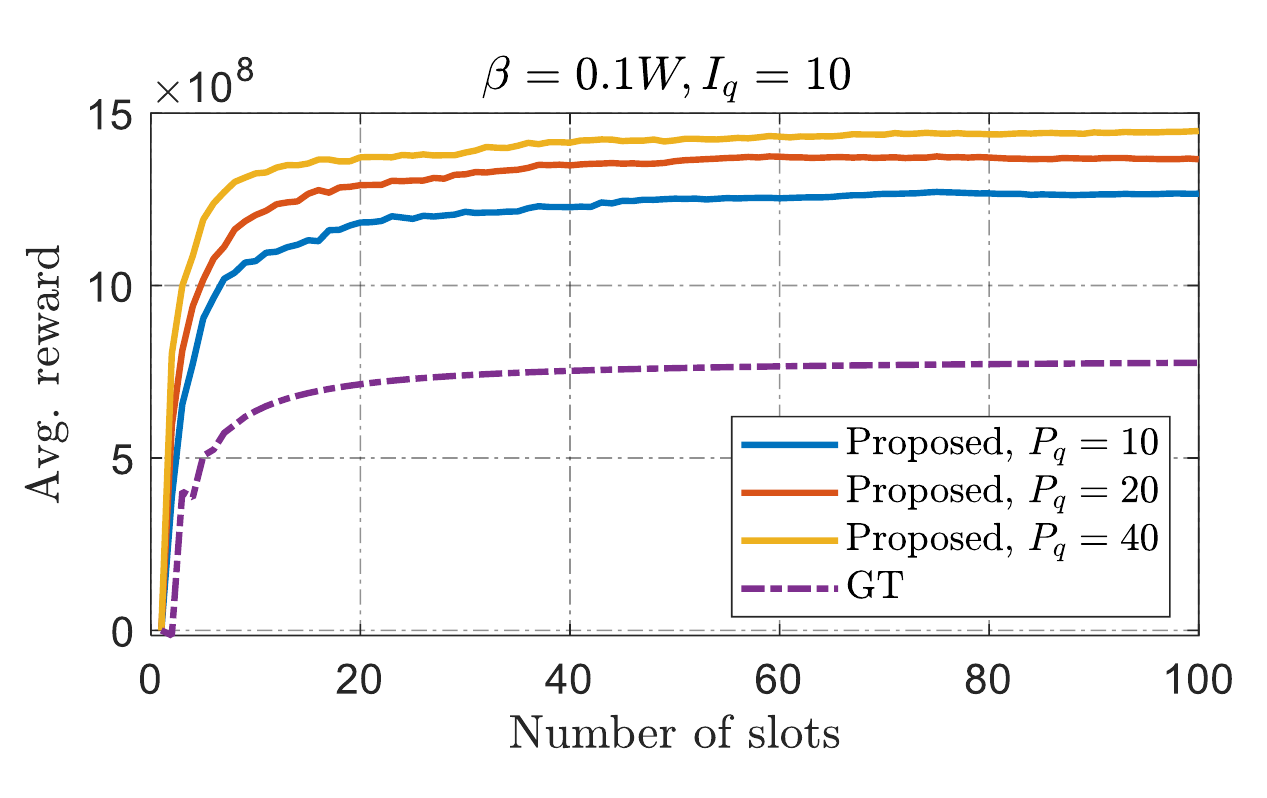}
        \vspace{-0.8cm}
        \caption{\small Effect of $P_q$ when $\beta =0.1W$, $I_q=10$.}
        \label{subfig 21}
    \end{subfigure}
    \begin{subfigure}[h]{0.4\textwidth}
        \includegraphics[width=\textwidth]{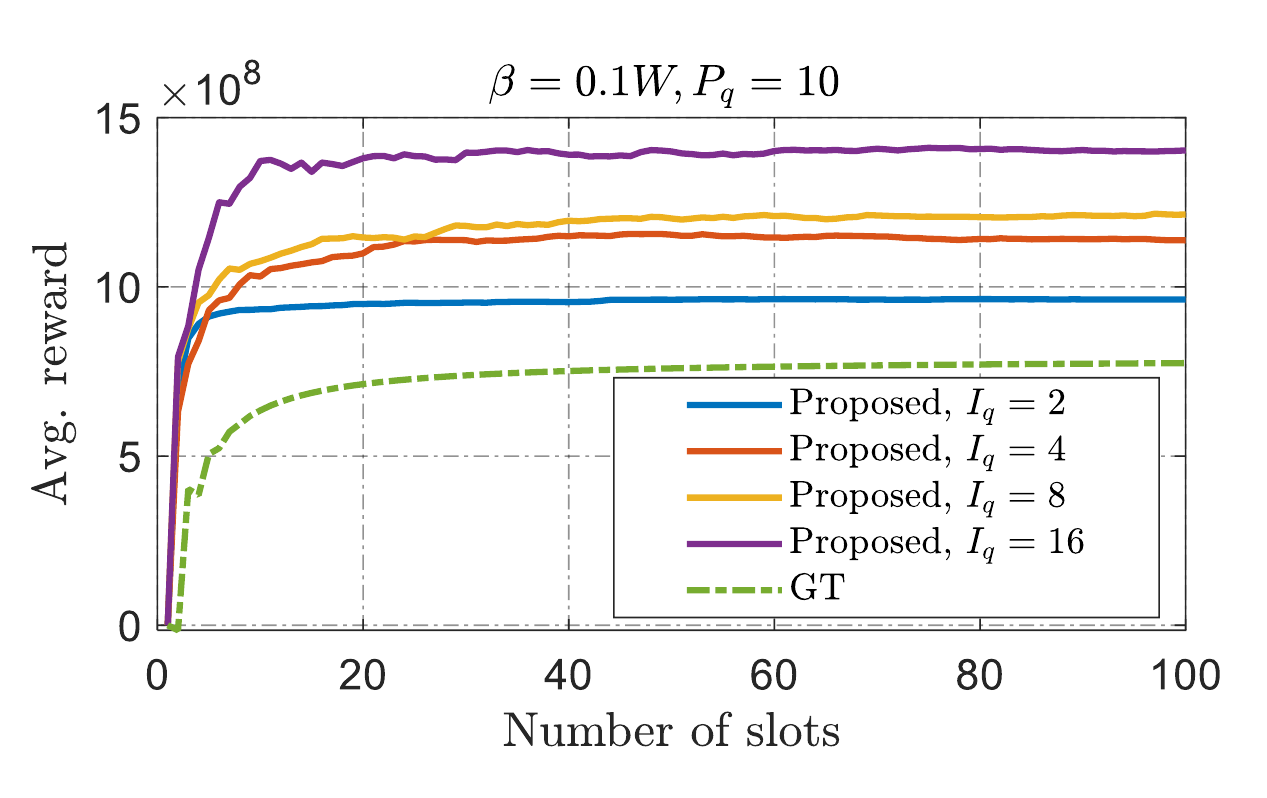}
        \vspace{-0.8cm}
        \caption{\small Effect of $I_q$ when $\beta =0.1W$, $P_q=10$.}
        \label{subfig 22}
    \end{subfigure}
    \vspace{-0.11cm}
\caption{\small Effect of $P_q$ and $I_q$ for different $\beta$. BSs have MSR of 20 dB and beamwidth $30^{\circ}$, UEs are omnidirectional.}
\label{fig: eff_of_Pq_n_Iq}
\end{figure*}

The BS antenna MSR and beamwidth are chosen to be 20 dB and $30^{\circ}$ respectively. The $1^{\rm st}$ UE of each BS is scheduled. This UE selection represents the behavior of the cell-edge UEs which usually suffer from strong interference from neighboring BSs. This phenomenon is even more prominent in ultra-dense small BS 5G cellular networks.  To verify the effect of $P_q$, we fix $I_q=10$ and let $P_q\in \{10,20,40\}$.  Figs.~\ref{subfig 11} and \ref{subfig 21} show the effect of $P_q$ for $\beta_i=0$ and $0.1W$ respectively. Each curve represents the average reward achieved up to the current slot, averaged over 50 independent trials each containing a set of i.i.d.  channel realizations. For both values of $\beta$, it can be seen that the proposed approach outperforms GT. For $\beta=0$, the proposed approach achieves 23\% to 39\% more average reward than GT in the $100^{\rm th}$ slot. For $\beta = 0.1W$, the proposed approach achieves 63\% to 87\% more average reward than GT. 
Moreover, the average reward increases as $P_q$ increases because larger $P_q$ provides more choices for power selection. To verify the effect of $I_q$, we fix $P_q=10$ and let $I_q\in \{2,4,8,16\}$. Figs.~\ref{subfig 12} and \ref{subfig 22} show the result. For both $\beta =0$ and $0.1W$, the achieved average reward of the proposed approach increases as $I_q$ increases. For $\beta=0$, when $I_q=2$, the proposed approach achieves a similar performance to GT. However, when $I_{q}=16$, there is a 33\% reward gain compared to GT. For $\beta=0.1W$, the proposed approach achieves 24\% to 80\% more reward than GT from $I_q=2$ to $I_q=16$. The effect of $I_q$ is expected because when there are more interference states for each agent, the decision making of each agent becomes more flexible and can cater to the specific interference condition according the agent's past experience.    

\subsubsection{Effect of beamwidth and MSR}
\begin{figure}[ht]
\centering
\includegraphics[width=0.42\textwidth]{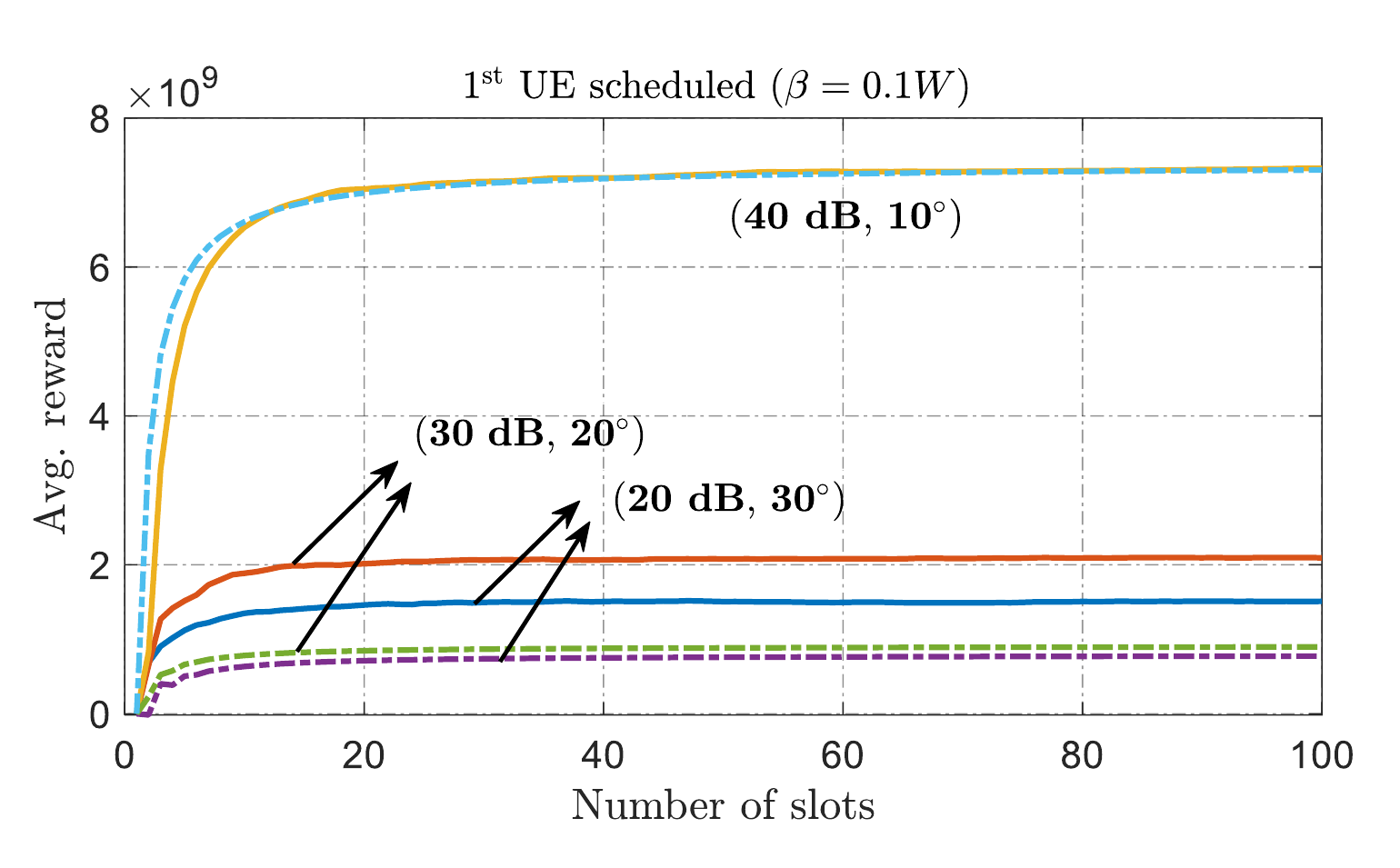}
\vspace{-0.3cm}
\caption{\small Proposed approach (solid lines) vs. game-based approach (dash lines) when the $1^{\rm st}$ UE of each BS is scheduled.}
\label{fig: eff_bw_msr_UE1}
\end{figure}

\begin{figure}[ht]
\centering
\includegraphics[width=0.42\textwidth]{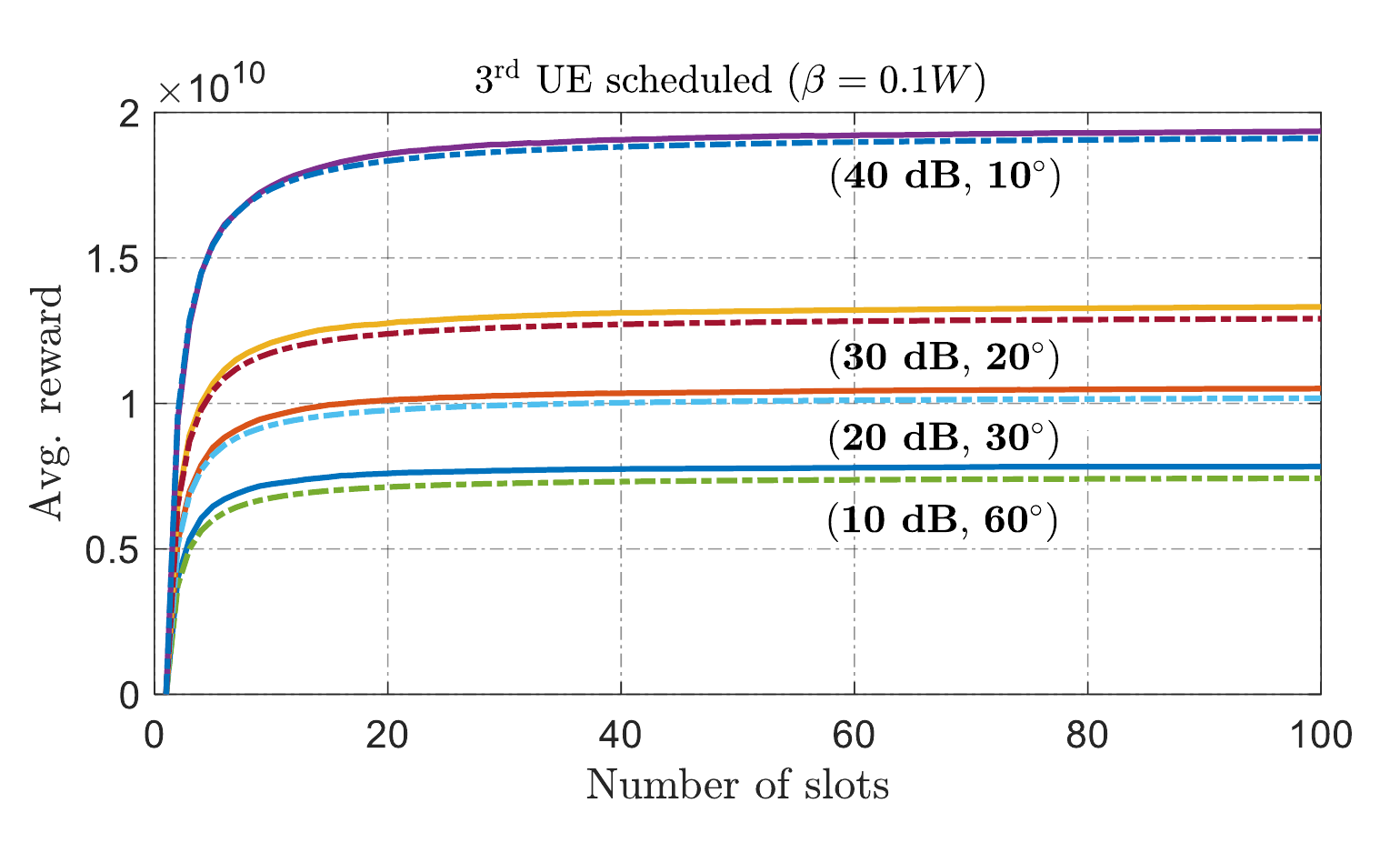}
\vspace{-0.3cm}
\caption{\small Proposed approach (solid lines) vs. game-based approach (dash lines) when the $3^{\rm rd}$ UE of each BS is scheduled.}
\label{fig: eff_bw_msr_UE3}
\end{figure}

The effect of beamwidth and MSR are shown in Fig.~\ref{fig: eff_bw_msr_UE1} and Fig.~\ref{fig: eff_bw_msr_UE3}. We fix $\beta=0.1W$. In Fig.~\ref{fig: eff_bw_msr_UE1}, the first UE of each BS is scheduled. These UEs represent the cell-edge UEs. We compare the performance of the proposed approach with GT under the BS antenna configurations $(20\,{\rm dB}, 30^{\circ})$, $(30\, {\rm dB}, 20^{\circ})$ and $(40 \,{\rm dB}, 10^{\circ})$. For the first two cases with BS beamwidth  $30^{\circ}$ and $20^{\circ}$, the proposed approach achieves 87\%  and 134\% more reward than GT. GT performs poorly in these cases by being greedy to choose the maximum power because there is beam overlapping which causes very strong interference to the non-target UEs due to high TX powers. This implies that the proposed approach has much better performance than GT in the interference-limited regime. 
However, when the beamwidth is further reduced to $10^{\circ}$, the proposed approach achieves a similar reward to GT. This is because in this case, BS beams are very sharp so they cause little
interference for non-target UEs. It has been shown by~\cite{zhang2020non} that when the interference level is very low, GT achieves near-optimal performance. Therefore, the proposed approach also achieves near-optimal performance in this case. 
Fig.~\ref{fig: eff_bw_msr_UE3} shows the case when the third UE of each BS is scheduled. Due to their separate locations, these UEs receive less interference and represent the cell-center UEs, which usually have high SINR. It can be seen that for any of the considered BS antenna configurations, the proposed approach outperforms GT by a small margin, and the margin diminishes as the beams become sharper (see the extreme case ($40\textrm{ dB},10^{\circ}$)). The reason for this competitive performance is that the interference level is relatively low because the scheduled UEs are sparsely distributed. This demonstrates that the proposed approach is at least as good as GT in the high SINR regime.

\section{Extensions} 
\subsection{Incorporation of the Lyapunov Optimization Framework}
\label{subsection: App to Lyapunov OPT}
One interesting aspect of the proposed approach is that the weights $\alpha,\beta$ can be automatically determined if we apply the Lyapunov optimization framework on top of the proposed power allocation  algorithm. More specifically, let us consider the following {\it utility maximization} problem
\begin{subequations}
\label{eq: sum utility maximization}
\begin{align}
 \max \quad &  \sum_{i\in \Mc}\sum_{j\in \Kc_i}  U(\bar X_{j,i})\label{eq 1} \\
\textrm{s.t.} \quad  
& \sum_{j\in\mathcal{K}_i}\bar{p}_{j,i} \le T_fp_i^{\rm avg},\quad \forall i, \label{eq 2}\\
&  p_{j_i,i}(k,n)\le p^{\rm max}_i,\quad \forall i,k,n,  \label{eq 3}
\end{align}
\end{subequations}
where $p_{j_i,i}(k,n)$ is the TX power of $\sfb_i$ in the $n^{\rm th}$ block of the $k^{\rm th}$ frame. Each $\sfb_i$ is subject to a long-term average and an instantaneous peak power constraint $p_i^{\rm avg}$ and $p_i^{\rm max}$ respectively. $\bar{p}_{j,i}$ represents the average power consumption of BS $i$ to UE $j$ in all frames.
$\bar{X}_{j,i}$ denotes the average 
number of received bits  by 
$\sfu_j$ in each frame and is referred to as the average  throughput in the following. $U(\cdot)$ represents 
the utility function, e.g., fairness function. 
Using the Lyapunov stochastic optimization framework~\cite{neely2010stochastic}, the above problem can be decomposed into two sub-problems to be solved in each frame, together with two virtual queues to enforce the average constraints. In particular, the first sub-problem aims to solve the  auxiliary variables $\gamma_{j,i}(k)$: 
\begin{subequations}
\label{Eq: sub-opt 1}
\begin{align}
\max \quad & \sum_{i\in\Mc}\sum_{j\in\mathcal{K}_i} V U({ \gamma_{j,i}(k)})  -{H_{j,i}(k)}\gamma_{j,i}(k)\label{Eq: sub-opt 11} \\
\textrm{s.t.} \quad & 0\le \gamma_{j,i}(k)\le T_fW\log\left(1+g_{j,i}^{\rm max}(k)p_i^{\rm max}\right),\; \forall i,j,k\label{Eq: sub-opt 12} 
\end{align}
\end{subequations}
where $V$ is a constant. $g_{j,i}^{\rm max}(k)\eqdef\max_{n}g_{j,i}(k,n)$ denotes the maximum equivalent channel gain in the $k^{\rm th}$ frame. $H_{j,i}(k)$ is the UE throughput queue which is  updated  by 
\begin{align}
\label{eq: queue H} 
& H_{j,i}(k+1) = \max\left\{  H_{j,i}(k)+\gamma_{j,i}(k)-X_{j,i}(k),\,0 \right\},\notag \\
& \qquad \qquad \qquad \qquad \qquad \qquad \qquad \;  \forall i\in\Mc,\forall j\in\Kc_i.
\end{align}
The second sub-problem aims to solve the TX powers $p_{j,i}(k,n)$:
\begin{subequations}
\label{Eq: sub-opt 2}
\begin{align}
\min  & \sum_{i\in\Mc}\sum_{j\in\mathcal{K}_i}     \left( \sum_{n\in[N_f]}     
\mathbb{E}\left[T_{j,i}^{d}(k,n){p_{j,i}(k,n)}\right]-T_fp_i^{\rm avg} \right) \notag\\
 &\qquad \qquad \qquad \qquad \quad \times Z_{i}(k)   -H_{j,i}(k)\widehat{X}_{j,i}(k)
\label{Eq: sub-opt 21} \\
\textrm{s.t.} \;& \;\;0\le  p_{j,i}(k,n)\le p_i^{\rm max},\quad \forall i,k,n \label{Eq: sub-opt 22}
\end{align}
\end{subequations}
where $${  \widehat{X}_{j,i}(k)\eqdef\sum\limits_{n=1}^{N_f}\EE\left[ T^{d}_{j,i}(k,n)W\log\left(1+\textrm{SINR}_{j,i}(k,n)\right)\right]}$$
denotes the expected throughput of $\sfu_{j}$ in the $k^{\rm th}$ frame. $T^{d}_{j,i}(k,n)$ denotes the data transmission time for UE $j$ by BS $i$ during block $n$ of frame $k$.  $Z_i(k)$ is the TX power queue which is updated by  
\begin{align}
\label{eq: queue Z} 
&Z_i(k+1)=\notag\\
&\max \bigg\{Z_i(k) + \sum_{j\in\mathcal{K}_i}\sum_{n\in[N_f]}  T_{j,i}^{d}(k,n)p_{j,i}(k,n) -T_fp_i^{ \rm avg},\,0 \bigg\},\notag\\
& \qquad \qquad \qquad \qquad\qquad \qquad \qquad \qquad \qquad \; \forall i\in \Mc.
\end{align}

\begin{figure}[ht] 
\centering
\includegraphics[width=0.44\textwidth]{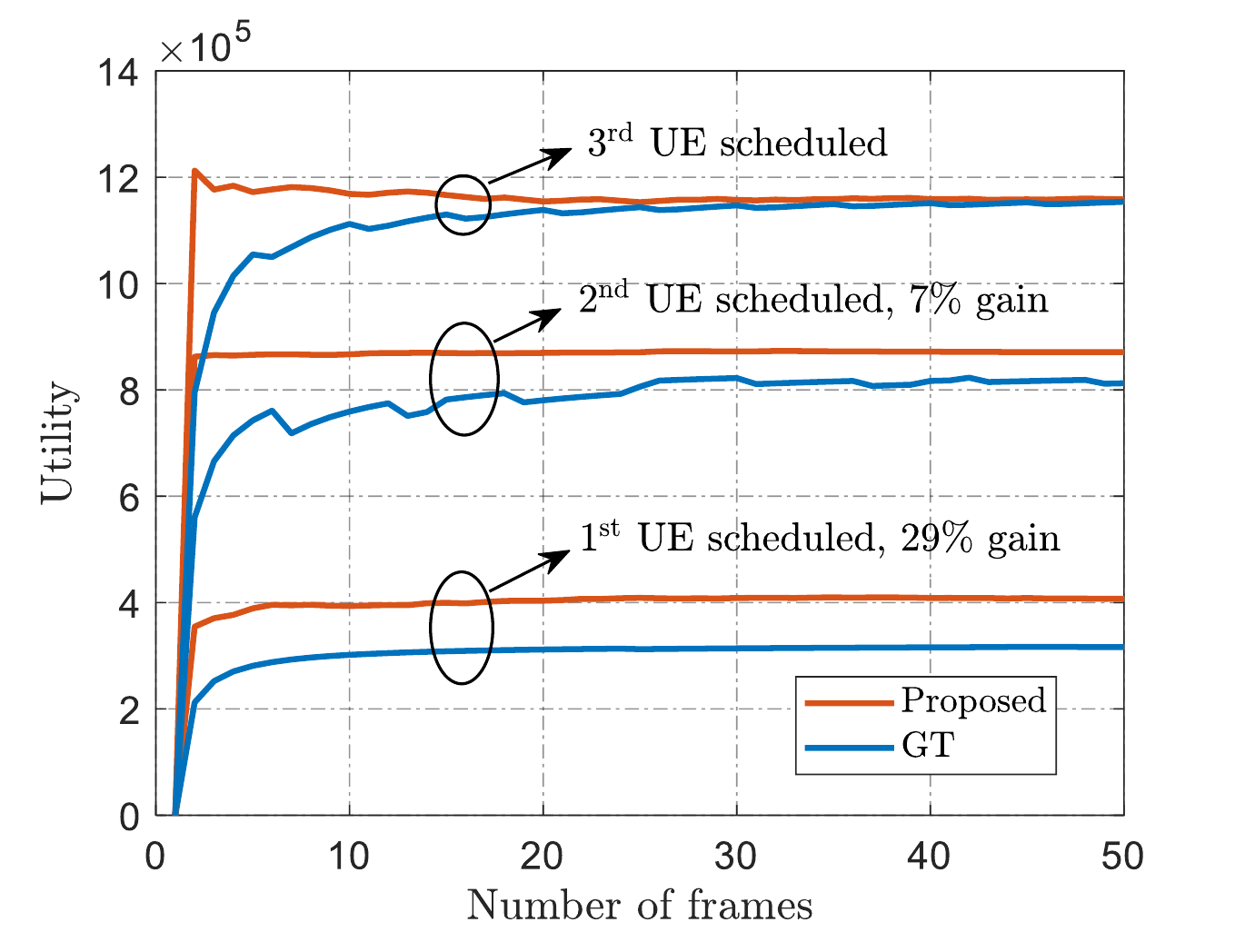}
\vspace{-0.2cm}
\caption{\small Proposed approach (red) vs. game-based approach (blue) when the Lyapunov framework is applied. }
\label{fig: LyOPT}
\end{figure}

Note that the objective of (\ref{Eq: sub-opt 21}) 
has the same form as the payoff function (\ref{eq: BS utility}) if we choose $\alpha_i=H_{j,i}(k)N_b$, $\beta_i = Z_i(k)N_b$. More specifically, given that $\sfu_{j_i}$ is scheduled, each $\sfb_i$ has an objective function $H_{j_i,i}(k)\widehat{X}_{j_i,i}(k,n) -Z_i(k) \EE\left[T_{j_i,i}^{d}(k,n){p_{j_i,i}(k,n)}\right]$ (the constant term $T_fp_i^{\rm avg}$ is omitted as it does not affect the optimal solution) to maximize in block $n$, where $\widehat{X}_{j,i_j}(k,n)$ is $\sfu_{j_i}$'s throughput in block $n$. By 
letting $\EE[T_{j_i,i}^d]=T_b$, i.e., the scheduled UE will be receiving data during the entire block, the objective becomes 
$\alpha_i T_sW\log(1+\textrm{SINR}_{j_i,i}(k,n))- \beta_i  T_sp_{j_i,i}(k,n)$. This objective can be optimized by maximizing the sum or average throughput in each of the $N_b$ slots in block $n$.
In this way, the proposed approach can be used to solve the second sub-problem (\ref{Eq: sub-opt 2}) in each block and in a distributed manner. It can be seen that the reward weights $\alpha_i,\beta_i$ are optimally determined by the virtual queues derived from the Lyapunov optimization framework. In \cite{zhang2020non}, the GT method (\ref{eq: GT update}) was used to solve the second sub-problem. Since we have shown that the proposed approach outperforms GT in a single block, it is expected to also achieve higher utility than GT when the Lyapunov framework is applied. Fig.~\ref{fig: LyOPT} shows the achieved utility when the $\alpha$-fair utility function $U(x) = x^{3/5}$ is used and under the same experiment setup as in Section \ref{sec: experiment}. BS beamwidth and MSR are chosen as $30^{\circ}$ and 20 dB while the UEs are ominidirectional.  
It can be seen that the proposed approach achieves 29\% more utility (at the $50^{\rm th}$ frame) than GT when the first UE of each BS is scheduled and 7\% more when the second UE is scheduled. For the cell-center UEs, i.e., the third UE of each BS, the proposed approach achieves a similar utility as GT but with a faster convergence. The queue values of $\sfb_1$ when the first UE  is scheduled is shown in Table~\ref{tab: 2}. It can be seen that $\beta_1/\alpha_1 = Z_1(k)/H_{1,1}(k)\approx 0,\forall k$. This mimics the behavior of the proposed power allocation algorithm when there is a very small penalty on power consumption.  
\begin{table}[htbp]
\caption{\small Virtual queue values corresponding to $\sfb_1$.}
\vspace{-0.2cm}
\begin{center}
\begin{tabular}{|c"c|c|c|c|c|}
\hline
Frame index $k$ & 10& 20&30 &40 & 50\\
 \thickhline 
$Z_1(k)$ & 0 & 0.24& 0&0 &0 \\
 \hline
${H_{1,1}(k)}/{10^9}$ &3.87& 0.14& 3.92 & 1.90 &0.11 \\
\hline
\end{tabular}
\end{center}
\label{tab: 2}
\end{table}

\subsection{Practical Consideration}
As discussed in Section~\ref{subsection: proposed approach}, the proposed approach adopts a per-BS storage complexity of $\Oc\left(\frac{KP_qI_q}{M} \right)$ and  a per-slot execution complexity of $\Oc\left(\max \{P_q,I_q\}\right)$. The storage complexity scales linearly with the number of UEs per BS and the execution complexity does not depend on the number of UEs. This demonstrates the scalability of the proposed approach However, to implement it on real-world cellular networks, 
there are still several practical considerations. First, in the proposed approach, the interference at the scheduled UE needs to be measured in each slot and then reported back to the associated BS. The measurements that can be configured at the UE for the relevant cellular systems, 5G and Beyond, has to be analyzed for an implementation.
Second, it is assumed in the proposed approach that the channels are block-fading and do not change within the duration of each scheduling block. Verification with real world cellular conditions will be appropriate for an implementation of this solution. 


\section{Conclusion}
We studied the problem of distributed beam scheduling and power allocation for non-cooperative mmWave networks. 
We presented a unified framework, with a flexible network payoff function definition, that can be used for  systematic performance evaluation and comparison of different scheduling methods. 
Furthermore, 
we propose a Q-learning-based approach using an independent agent modeling where each BS can adaptively control its transmit power for different interference situations based on its experience and active exploration of non-greedy actions.
Experiments showed that the proposed approach
outperforms the non-cooperative game-based approach in the sense that they achieve similar performance in the high SINR regime but the proposed approach beats the game-based approach by a large margin in the interference-limited regime. 
In addition, the proposed approach can be integrated into the Lyapunonv stochastic optimization framework for the purpose of network utility maximization. In this case, the weights in the reward function are automatically and optimally determined by the virtual queues. 



\bibliographystyle{IEEEtran}
\bibliography{references_d2d}
\end{document}